\documentclass[pra,twocolumn,notitlepage,aps,superscriptaddress,footinbib]{revtex4-1}
\usepackage{graphicx}
\usepackage{xcolor}
\usepackage{amsmath}
\usepackage{bm}
\usepackage{tabularx}
\usepackage[percent]{overpic}
\usepackage{ulem}
\usepackage{soul}

\newcommand{\kp}{\ensuremath{{\bm k}\cdot{\bm p}}}
\newcommand{\ignore}[1]{}

\allowdisplaybreaks

\begin{document}

\title{Non-symmorphic symmetry and field-driven odd-parity pairing in
  CeRh$_2$As$_2$} 

\author{D. C. Cavanagh}
\affiliation{Department of Physics, University of Otago, P.O. Box 56, Dunedin 9054, New Zealand}
\author{T. Shishidou} 
\affiliation{Department of Physics, University of Wisconsin, Milwaukee, Wisconsin 53201, USA}
\author{M. Weinert} 
\affiliation{Department of Physics, University of Wisconsin, Milwaukee, Wisconsin 53201, USA}
\author{P. M. R. Brydon}
\email{philip.brydon@otago.ac.nz} 
\affiliation{Department of Physics and MacDiarmid Institute for Advanced Materials and Nanotechnology,
University of Otago, P.O. Box 56, Dunedin 9054, New Zealand}

\author{Daniel F. Agterberg}
 \email{agterber@uwm.edu}
\affiliation{Department of Physics, University of Wisconsin, Milwaukee, Wisconsin 53201, USA}

\begin{abstract}
Recently, evidence has emerged for a field-induced even- to odd-parity
superconducting phase transition in CeRh$_2$As$_2$ [S. Khim {\it et al.},
Science {\bf 373} 1012 (2021)]. Here we argue that the P4/nmm non-symmorphic crystal structure of CeRh$_2$As$_2$ plays a key role in enabling this transition by ensuring large spin-orbit interactions near the
Brillouin zone boundaries, which naturally leads to the required near-degeneracy
of the even- and odd-parity channels. We further comment on the
relevance of our theory to FeSe, which crystallizes in the same
structure. 
\end{abstract}

\maketitle

{\it Introduction---}The discovery of a transition between two distinct superconducting
phases at high magnetic fields in CeRh$_2$As$_2$~\cite{Ce122_arxiv} has generated great
interest~\cite{Schertenleib_2021,Ptok_2021,Moeckli_2021,Skurativska_2021,Nogaki_2021}. Due
to the immense upper critical field, this has been widely interpreted
as a transition between even- and odd-parity pairing states.
Creating
  odd-parity superconductors is a central goal of quantum
  materials science  as they can host non-trivial
  topological phenomena~\cite{Sato_2017}.  The putative field-induced
    transition in CeRh$_2$As$_2$ offers a straightforward
    route to a bulk odd-parity state. As such, it is of great
    importance to clarify the physics responsible for its remarkable
    phase diagram.
    
The even to odd parity transition is enabled by a Rashba-like
spin-orbit coupling (SOC) that exists on an inversion ($I$) symmetry
breaking sublattice of atoms. $I$ symmetry transforms one sublattice
to the other, with opposite signs for SOC, ensuring
the Hamiltonian satisfies a global $I$ symmetry. However, the even to
odd parity transition is also suppressed by hopping between the two
sublattices, and so 
 the SOC should be larger than this inter-sublattice hopping for this
transition to occur.
It is unclear 
if this condition can be realized in bulk crystals:   Indeed, a
  relatively strong Rashba-like spin texture has been observed
  in bilayer cuprate Bi2212 \cite{Gotlieb_2018}, but
  there is no evidence of a field-induced odd-parity state.
  Since 
    superconductors with this sublattice structure are not uncommon,
    the rarity of the 
    field-induced transition suggests that additional physics is
    necessary to  explain 
 the phase diagram of CeRh$_2$As$_2$. 
    
 Here  we show that the non-symmorphic (NS) structure of CeRh$_2$As$_2$
 allows the Rashba-like SOC to be larger than the inter-sublattice
 hopping, providing an explanation for why this transition is
 observed. In particular, we show that the NS structure
 ensures that the SOC energy scale is asymptotically larger than that
 of the inter-sublattice hopping near the Brillouin zone edges. 
 Provided that a Fermi surface with sufficiently large density of
 states  (DOS)  exists near the zone edge, the field-induced even- to
 odd-parity transition can appear at the relatively high temperature
 seen 
in CeRh$_2$As$_2$.

The manuscript is organized
as follows: First a general argument is given that
 on the Brillouin zone edge 
  arbitrary superpositions of Kramers degenerate
states have the same spin polarization direction,  in contrast to Kramers degenerate states at the Brillouin zone center. This remarkable feature
reflects the dominance of SOC near the zone edges.  We confirm this by examining a
$\kp$ theory valid near the zone edge and contrasting
it with one valid near the zone center. This explicitly
reveals that the SOC is asymptotically smaller than the
inter-sublattice hopping  near the zone center, but is
asymptotically larger near the zone edge. 
Considering superconductivity originating from 
an intra-sublattice pairing instability, the dominance of the
  SOC at the zone edge allows us to
qualitatively
reproduce the magnetic field-temperature phase diagram of
CeRh$_2$As$_2$, provided that the contribution to
   the DOS  from the Fermi surfaces near the
zone edges is sufficiently large. 
Density functional calculations for CeRh$_2$As$_2$ reveal that this can be
the case 
if electron correlations are included. Finally, since our analysis shows that exotic physics due to strong
SOC can be expected generically in NS
superconductors,  we discuss an additional application of our theory to NS FeSe.

{\it Non-symmorphic symmetry and spin texture---} In contrast to Kramers degenerate band states at the Brillouin zone center, the NS
  structure of CeRh$_2$As$_2$ implies that the two-fold Kramers degenerate band states at the zone edge exhibit the same
 spin polarization direction. 
To show this, we consider the set of symmetries that keep  momenta lying in the zone-center plane ${\bm k}_c=(0,k_y,k_z)$ and the zone-edge plane  ${\bm
  k}_e=(\pi,k_y,k_z)$ unchanged. These include: $M_x$, a mirror
reflection through the $\hat{x}$ direction; $T\tilde{I}$,  where $T$
is time-reversal symmetry and
$\tilde{I}=\{I|\frac{1}{2},\frac{1}{2},0\}$; and their product  $T\tilde{I}M_x$. Since $(T\tilde{I})^2=-1$, these states exhibit a two-fold Kramers degeneracy denoted as $|{\bm k}_{\nu=e,c},\pm\rangle \equiv |\nu,\pm\rangle$.
These two-fold degenerate
eigenstates are also eigenstates of $M_x$, and since $M_x^2=-1$,
$M_x|\nu,\pm\rangle=e_{\nu,\pm}|\nu,\pm\rangle$ where the $e_{\nu,\pm}$ are purely imaginary. 
 From  the non-symmorphicity and the general result that $T$
commutes with spatial 
symmetries, we find $T\tilde{I} M_x=\{E|100\}M_x T\tilde{I}$, where
$\{E|100\}$ is an in-plane translation vector. Importantly,
$\{E|100\}$ becomes 1 for $k_x=0$, and $-1$ for $k_x=\pi$. Hence,
for the zone center plane $M_x (T\tilde{I}|c,\pm\rangle)=-e_{c,\pm} (T\tilde{I}|c,\pm\rangle)$,
while for the zone-edge plane $M_x (T\tilde{I}|e,\pm\rangle)=e_{e,\pm}
(T\tilde{I}|e,\pm\rangle)$. We thus conclude that Kramers degenerate partners at the
zone center plane have {\it opposite} eigenvalues  with respect to $M_x$,
while those at the zone boundary have the {\it same} eigenvalue  with
respect to $M_x$.  

Since the spin operators $S_y$ and $S_z$ are odd under
$M_x$,  they will have non-zero matrix elements at the
zone-center plane, but all their matrix elements are zero at the zone-edge
plane~\footnote{A proof is as follows: The matrix elements of
  the $j$ spin operator with respect to the pseudospin partners are
$\protect\langle \nu,m|S_j|\nu,m' \protect\rangle =
\protect\langle \nu,
m|\tilde{M}_x^{-1}\tilde{M}_x S_j \tilde{M}^{-1}_x\tilde{M}_x|\nu,m'\protect\rangle
= e_{\nu,m}^* e_{\nu,m'} (-1)^{j}\protect\langle \nu,m|S_j|\nu,m'\protect\rangle$, where $(-1)^j=-1$ ($1$) 
for $j = z$, $y$ ($x$), and $e_{\nu,m}$ is the eigenvalue of
$\tilde{M}_x$. Since the states at the zone boundary have the same
eigenvalues, we conclude that all the matrix elements for
the $z$ and $y$ spin operators are vanishing.}. That is, at the zone
edges, all eigenstates have their spins polarized along the $\pm \hat{x}$
direction.  A similar argument can be made for the ${\bm
  k}=(k_x,\pi,k_z)$ plane, where we find that the states are polarized
along the $\pm\hat{y}$ direction.
This implies that the 
states in the zone-edge
plane cannot couple to a $c$-axis field, while those in the zone
center plane can.   As seen below, it is this key
difference that enhances the 
$c$-axis Pauli limiting fields and stabilizes a field-induced even- to
odd-parity transition for Fermi surfaces 
near the zone edges
relative to those 
near the zone center.

{\it $\kp$ theories---} 
To quantify the difference between zone-center
and zone-edge Fermi surfaces on superconductivity, we  construct $\kp$ theories valid
near the $\Gamma$ point and the M-A Dirac line.  (Our
results  for the M-A Dirac line hold more generally for Fermi surfaces near the zone-edge.) For the $\Gamma$
point, it suffices to take the small ${\bm k}$ limit of the tight
binding theory presented in Ref.~\cite{Ce122_arxiv}.  To develop the
$\kp$ theory near the M-A Dirac line \cite{Young_2012} we use the representations as
given on the Bilbao crystallographic server~\cite{bilbao1, *bilbao2}. 

In both $\kp$ theories the Hamiltonian has the structure
\begin{align}
H_0=&\epsilon_{00, {\bm k}}\tau_0\sigma_0+\epsilon_{x0, {\bm
      k}}\tau_x\sigma_0+\epsilon_{y0, {\bm k}}\tau_y\sigma_0 \notag \\
& +\epsilon_{zx, {\bm k}}\tau_z\sigma_x+\epsilon_{zy, {\bm
  k}}\tau_z\sigma_y+\epsilon_{zz, {\bm k}}\tau_z\sigma_z. \label{eq:H0}
\end{align}
The $\tau_i$  Pauli matrices encode the sublattice basis
composed of two states that are transformed into each other under 
 inversion 
(e.g., a Ce site basis). The $\sigma_i$ Pauli matrices encode the spin basis.
The first line of Eq.~(\ref{eq:H0}) describes spin-independent intra-
and  inter-subllatice
hopping processes, whereas the second line includes the SOC terms. $I$ symmetry is given by the operator
$\tau_x\sigma_0$ at the $\Gamma$, M, and A points. Consequently,
$\epsilon_{00,{\bm k}}$ and $\epsilon_{x0,{\bm k}}$ are even in
  momentum $\bm k$, while the other coefficients are odd.  Eq.~(\ref{eq:H0}) has the same form as a minimal Hamiltonian for a
locally non-centrosymmetric
material~\cite{FuBerg_2010,Youn_2012,Yanase_UPt3_2016,Xie_WTe2_2020,Shishidou_UTe2_2021}. 
The Hamiltonian possesses two doubly-degenerate eigenvalues $\epsilon_{00,{\bm k}}\pm \tilde{\epsilon}_{\bm k}$ where
\begin{equation}
\tilde{\epsilon}_{\bm k}=\sqrt{\epsilon_{x0,{\bm
             k}}^2+\epsilon_{y0,{\bm k}}^2+\epsilon_{zx,{\bm
             k}}^2+\epsilon_{zy,{\bm k}}^2+\epsilon_{zz,{\bm k}}^2}.
\end{equation}
\noindent It is convenient to label the two degenerate states in
each band by a pseudospin index. Our choice of
pseudospin basis is presented in the SM~\footnote{See the Supplemental Material at
  .... for details of the DFT calculations, the pseudospin
  basis, and the calculation of the phase diagram. This includes Ref.s~\cite{Mike_FLAPW_2009,PBE,KAWAMURA2019197,Momma2011,varma76,npjqm2018,Fu_MCBB}.}.

In Table~\ref{tab:kp} we give the momentum dependence of the coefficients
$\epsilon_{\mu\nu,{\bm k}}$. Along the M-A line  we expand radially
from the line, i.e., ${\bm
  k}=(\pi,\pi,k_z)+(k_x,k_y,0)$ and expand in $k_x$ and $k_y$. We do
not give the form of $\epsilon_{00,{\bm k}}$ since this  term does not
play an essential role in the physics, and also only keep the lowest
non-zero power of  $k_\nu$ in the coefficient of each
  $\tau_i\sigma_j$ matrix. The $\kp$ theories reveal several  remarkable features of the
electronic structure: i) The
$\epsilon_{zz,{\bm k}}$ SOC is parametrically smaller
than the Rashba-like SOC terms $\epsilon_{zx,{\bm k}}$ and
$\epsilon_{zy,{\bm k}}$ (and will henceforth be ignored); ii) Near the M-A Dirac line when $k_x=0$, only the coefficient of $\tau_z\sigma_x$ is non-zero,  a consequence of the NS spin texture presented above; iii) The NS symmetry requires that all
coefficients vanish at the M-A Dirac line, and hence the energy bands are
four-fold degenerate here. 

Importantly, the Rashba SOC terms vanish
asymptotically more slowly than the inter-sublattice hopping as the M-A Dirac line is approached. This is
reflected in the divergence of the ratio $\tilde{\alpha}_{\bm k} = \sqrt{(\epsilon_{zx,{\bm k}}^2
  + \epsilon_{zy,{\bm  
      k}}^2)/(\epsilon_{x0,{\bm k}}^2
  + \epsilon_{y0,{\bm  
      k}}^2)}$ as one approaches the Dirac line. In contrast, only the inter-sublattice hopping $\epsilon_{x0,{\bm
    k}}$ can be nonzero at the $\Gamma$ point, which  implies
  that the ratio $\tilde{\alpha}_{\bm k}$ vanishes at the zone
  center.  As we shall see,  $\tilde{\alpha}_{\bm k}$ plays
a key role in our theory.

{\it Zeeman response---}We include a Zeeman field by adding the term $H_Z = g \mu_B\tau_0
\vec{\sigma}\cdot\vec{H}$ to the Hamiltonian~Eq.~(\ref{eq:H0}).
 Expressed in the band-pseudospin basis, $H_Z$ typically has both
interband and intraband components. The former are not important in the $\tilde{\epsilon}_{\bm
  k}\gg g\mu|\vec{H}|$ limit; in contrast, the latter lifts
the pseudospin
degeneracy, acting like an effective pseudospin Zeeman field, which we obtain
by projecting
$\tau_0\vec{\sigma}$ onto the pseudospin basis, $\tau_0\sigma_\mu \rightarrow
\vec{\gamma}^\mu_{\bm k}\cdot\vec{s}$. 
For our choice of pseudospin basis, Zeeman fields parallel
(perpendicular) to the $c$-axis produce pseudospin fields that are
also parallel (perpendicular) to the $c$-axis; explicit
expressions for the effective $g$-factors $\vec{\gamma}^\mu_{\bm k}$
are given in the SM~\cite{Note2}.  
Moreover, the magnitude of $\vec{\gamma}^\mu_{\bm k}$ is
basis-independent and given by
\begin{equation}
|\vec{\gamma}^{\mu}_{\bm k}|^2=\hat{\epsilon}_{x0,{\bm
                              k}}^2+\hat{\epsilon}_{y0,{\bm
                              k}}^2+\hat{\epsilon}_{z\mu,{\bm k}}^2 
\end{equation}
where $\hat{\epsilon}_{\mu\nu,{\bm k}} = \epsilon_{\mu\nu,{\bm
    k}}/\tilde{\epsilon}_{\bm k}$.  For a $c$-axis field, the
pseudospin splitting is controlled by the ratio $\tilde{\alpha}_{\bm
  k}$ as  $|\vec{\gamma}^z_{\bm k}| = (1 + \tilde{\alpha}_{\bm
  k}^2)^{-1/2}$. Our $\kp$ theory therefore 
shows that the 
pseudospin  splitting is maximal near 
the $\Gamma$
point, but vanishes as we approach the M-A Dirac line. This reflects
the in-plane spin polarization of the band states near the zone
edge required by the NS symmetry as discussed above, and implies that the
effective $g$-factor  vanishes 
on the zone boundary ($|\vec{\gamma}^z_{\bm k}|=0$). 

\begin{table}
\begin{tabular}{ccc}
\hline
 & $\Gamma$ point& M-A Dirac line\\\hline\hline

$\epsilon_{x0,{\bm k}}$ & $g_{x0}$& $m^{(e)}_{x0} k_xk_y$\\
$\epsilon_{y0,{\bm k}}$ &  $g_{y0}k_z$  & $m^{(o)}_{y0} k_xk_y$\\
$\epsilon_{zx,{\bm k}}$  & $g_{zx} k_y$ & $m^{(e)}_{zx} k_y$\\
$\epsilon_{zy,{\bm k}}$  & $-g_{zx} k_x$ & $-m^{(e)}_{zx} k_x$\\
$\epsilon_{zz,{\bm k}}$  & $g_{zz} k_zk_xk_y(k_x^2-k_y^2)$ &$m^{(o)}_{zz}                                                                                 k_xk_y(k_x^2-k_y^2)$\\\hline
$\tilde{\alpha}_{\bm k}^2 $ & $\frac{g_{xz}^2(k_x^2+k_y^2)}{g_{x0}^2}\rightarrow 0$ & $\frac{m_{zx}^{(e)2}(k_x^2+k_y^2)}{(m_{x0}^{(e)2}+m_{y0}^{(o)2})k_x^2k_y^2}\rightarrow\infty $\\\hline\hline
\end{tabular}
\caption{Form of the nontrivial terms in Eq.~(\ref{eq:H0}) near the $\Gamma$ point and along the  M-A Dirac line. The
  expansion coefficients $m_{ij}^{(e)}$ are 
  functions of $k_z$, with the $(e)$ and $(o)$ superscripts indicating
  that these are non-vanishing or vanishing at the $M$ and
  $A$ points. The last row gives $\tilde{\alpha}_{\bm k}^2$,
  characterizing the ratio of the SOC to the
  inter-sublattice hopping, and the limiting values as ${\bm k}\rightarrow 0$.} \label{tab:kp}
\end{table}

{\it Superconductivity---}In the standard scenario for the field-induced transition in
  locally $I$-symmetry broken 
  superconductors~\cite{Fischer_2011,Yoshida_2012,Yoshida_2014,Sigrist_2014},
  the dominant interaction pairs electrons on the same sublattice in a
  spin singlet. Since the sublattices are swapped by $I$, this
  generates both even- and odd-parity states, corresponding to equal
  and opposite  signs of the
  pairing potential on each sublattice, 
  respectively. We refer to these two possibilities as the
  uniform and staggered states.  The sign difference can be
    readily encoded in the $\tau$-dependence of the pairing potential,
    which for the  
  uniform (staggered) state is $f_{\bm k}\Delta \tau_0i\sigma_y$
  ($f_{\bm k}\Delta\tau_zi\sigma_y$), where $f_{\bm k}$ is an
  even-parity form factor.

In the pseudospin basis, the uniform and staggered pairing
potentials are
\begin{eqnarray}
 \Delta f_{\bm k} \tau_0i\sigma_y& \rightarrow &\Delta f_{\bm k} is_y \label{eq:projunif}\\
 \Delta f_{\bm k} \tau_zi\sigma_y & \rightarrow &  \pm \left(\hat{\epsilon}_{zx,{\bm k}}
                                        s_x+\hat{\epsilon}_{zy,{\bm
                                                  k}}s_y\right) \Delta
                                                  f_{\bm k} is_y\,. \label{eq:projstag}
\end{eqnarray}
The odd-parity staggered state is transformed into a helical
pseudospin-triplet state, with reduced gap magnitude $\sqrt{\hat{\epsilon}_{zx,{\bm k}}^2 + \hat{\epsilon}_{zy,{\bm
      k}}^2}|\Delta f_{\bm k}|$ and opposite sign in each band.  The
reduced gap magnitude of the 
  staggered state is due to interband  pairing,
implying that this state has a lower transition temperature ($T_c$) than the  uniform
state. In 
the weak-coupling limit the $T_c$ of the
  staggered state is determined by an effective
  coupling constant which is smaller than that of the
  uniform state by  $\langle \hat{\epsilon}_{zx,{\bm k}}^2
  + \hat{\epsilon}_{zy,{\bm  
      k}}^2 \rangle_{\text{FS}} = \langle \tilde{\alpha}_{\bm k}^2/(1
  + \tilde{\alpha}_{\bm k}^2)\rangle_{\text{FS}}$
where the average is taken over the Fermi
  surface~\cite{Ramires2018}.
Due
to the exponential sensitivity of $T_c$ on the
coupling constant, the ratio $\tilde{\alpha}_{\bm k}$ must be larger
than unity for $T_c$ of the staggered and uniform
states to be comparable.

 The projection onto the pseudospin basis reveals the essential physics
 of the field-induced transition. Since the same interaction
 mediates pairing in both channels, the
 generically smaller gap opened by the staggered state implies that it
 has the lower $T_c$ at zero field. However, whereas the
 uniform state is Pauli limited (albeit with an
  enhanced upper critical field due to the reduced effective $g$
  factor~\cite{Xie_WTe2_2020}), the staggered state is not
  Pauli limited for a $c$-axis field, since the effective
  pseudospin Zeeman field
 is perpendicular to the
  $\vec{d}$-vector of the pseudospin triplet state. Thus, a
  field-induced transition occurs when a
  $c$-axis field suppresses the uniform
  state below the $T_c$ of the staggered state.

The key parameter that underlies both the $T_c$ of the staggered state and the response of the uniform state to $c$-axis fields is $\tilde{\alpha}_{\bm k}$. Crucially, our $\kp$ analysis shows that $\tilde{\alpha}_{\bm k}$ 
strongly varies 
across the
Brillouin zone in CeRh$_2$As$_2$ due to the NS crystal symmetry. 
In particular, although it vanishes upon approaching the $\Gamma$
point, $\tilde{\alpha}_{\bm k}$ diverges towards the M-A Dirac line due to the vanishing
inter-sublattice terms, as indicated in
Table~\ref{tab:kp}. More generally, it diverges on the Brillouin zone edges. Thus, large values of $\tilde{\alpha}_{\bm k}$ are
generically expected for Fermi surfaces sufficiently near the zone edge. In CeRh$_2$As$_2$, the field-induced
  transition occurs at $T_{c,t}\approx 0.7 T_{c,0}$, where $T_{c,0}$ is the
  zero-field transition temperature, implying that $\tilde{\alpha}_{\bm
    k}\approx 3.5$ at the Fermi energy. Our
  theory shows that such ratios are possible 
  if states near the Brillouin zone edge make a significant contribution to the
  DOS at the Fermi energy.  
 Previous theoretical studies of CeRh$_2$As$_2$ have assumed 
  Fermi surfaces near the $\Gamma$ point, where the 
  enhancement of $\tilde{\alpha}_{\bm k}$ due to the NS
  symmetry is not
  apparent~\cite{Moeckli_2021,Skurativska_2021,Nogaki_2021}; as in similar
 treatments of symmorphic
 lattices~\cite{Fischer_2011,Yoshida_2012,Yoshida_2014,Sigrist_2014},
 these theories require  an unexpectedly large SOC
 strength to explain the field-induced transition.

\begin{figure}
\includegraphics[width=0.95\columnwidth]{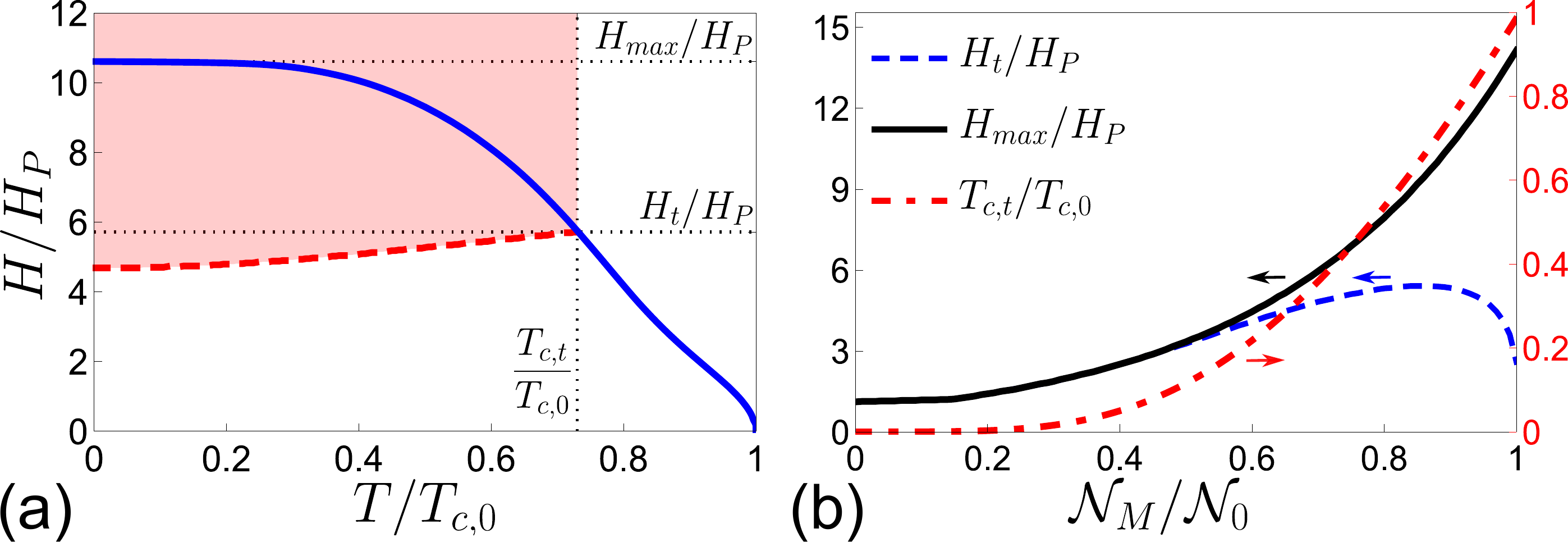}
\caption{ (a) Phase diagram
for $\mathcal{N}_{M}=0.9\mathcal{N}_{0}$: the blue line gives the upper critical field
of the uniform state, and the red dashed line gives the boundary of
the staggered state. The staggered state is realized in the shaded region.   (b) The dependence  of the upper critical field of
the uniform state ($H_{\text{max}}$, black solid line) and the critical field ($H_t$, blue dashed line)
and temperature ($T_{c,t}$ red dot-dashed line) at which the field-induced transition
occurs as a function of the contribution of the density of states at
the M point to the total density of states.  The field strengths are
expressed in terms of the Pauli limiting field $H_P\approx 1.25 k_BT_{c,0}$.}\label{fig:TwoPocket}
\end{figure}

It is instructive to contrast our results with previous results in $I$-symmetric 2D
Ising superconductors
\cite{Nakamura_2017,Wang_2019,Falson_2020}  and a toy model of a 1D NS zig-zag chain \cite{Sumita_2016}. In the Ising systems, a
 symmetry-required divergence
of the ratio $\tilde{\alpha}_{\bm k}$ occurs for band representations
with angular momentum $j_z=\pm 3/2$ at  certain points in the 2D
Brillouin zone, which strongly enhances the Pauli limit field for
in-plane fields. 
Our result is more
general, however, as the divergent $\tilde{\alpha}_{\bm k}$ occurs on a 2D
manifold of the 3D Brillouin zone, and holds for all band
representations.  In the zig-zag chain, the stability of an odd-parity state similar to that discussed here is found to be enhanced when the 1D FS is near the zone edge \cite{Sumita_2016}. Although the corresponding ratio $\tilde{\alpha}_k$ does take a maximum at the zone-edge, it does not diverge as in our model. Consequently, the NS spin texture mechanism we examine
is a more general route to enhancing the effect of SOC. 


%
{\it Two-pocket model---}While 
Fermi surfaces near the zone edge favor a field-induced transition, it
is likely that they will appear
together with  other Fermi surfaces  near the zone center where
the parameter $\tilde{\alpha}_{\bm k}$ is small. To examine the
sensitivity of our theory to the presence of these additional Fermi
surfaces,  we  consider a model of CeRh$_2$As$_2$ with two cylindrical Fermi 
pockets centered on the $\Gamma$-Z and M-A Dirac lines, fixing
$|\vec{\gamma}^z_{\bm k}| = 0.9$ and  $0.1 |\hat{k}_x\hat{k}_y|$ on the two
Fermi surfaces, corresponding to small and large values of
$\tilde{\alpha}_{\bm k}$, respectively. The momentum-dependence of the
effective $g$-factor near the M-A Dirac line reflects the NS
symmetry-enforced spin texture at the zone boundary. Assuming  an
  intrasublattice pairing interaction,
we use standard 
techniques to construct the field-temperature phase diagram,
see the SM for details~\cite{Note2}. For simplicity we
assume an $s$-wave form factor, i.e. $f_{\bm k}=1$, but our results
are robust to other choices.

 In Fig.~\ref{fig:TwoPocket}(a), we present a phase diagram  which
qualitatively agrees with that observed in CeRh$_2$As$_2$.
Since we only consider the Zeeman effect, the upper critical
field of the staggered state is infinite,  and so the right-most
  boundary of the staggered state is
vertical; including orbital effects will give a finite upper critical
field~\cite{Schertenleib_2021}, but does not qualitatively alter our
theory. Fig.~\ref{fig:TwoPocket}(a) was found by setting the M-A
pocket DOS at $90$\% the total DOS.  In 
Fig.~\ref{fig:TwoPocket}(b) we examine the consequences of varying
this M-A pocket DOS 
for the upper critical field of the
uniform state, and the field strength and temperature at which the
transition into the staggered state occurs.
The field-induced transition  is strongly enhanced as the
contribution of the M-A pocket to the DOS increases, with the even-
and odd-parity states near-degenerate when this is the only Fermi
surface.  Importantly, the field-induced transition occurs at an
observable temperature
$T_{c,t}>0.1T_{c0}$ if the M-A pocket makes up at least half of the
DOS. 

{\it DFT results---}DFT calculations and analysis were carried out to explore 
the possibility that the Fermi surface of CeRh$_2$As$_2$ contains regions near the
zone-edge and to verify that the states at the zone-edge exhibit the
spin polarization found above. 
As shown in the SM~\cite{Note2}, the Fermi surface (shown in Fig.~S1a) predicted by the
DFT bands
 consists of 
four pockets about the A point that do not intersect the zone edge, and
portions about the $\Gamma$-Z line, representing $\sim$53\% of the
DOS, 
 in agreement with
\cite{Nogaki_2021}. 
This Fermi surface is unlikely to be consistent
with the observed odd-parity state.


\begin{figure}
\includegraphics[width=0.9\columnwidth]{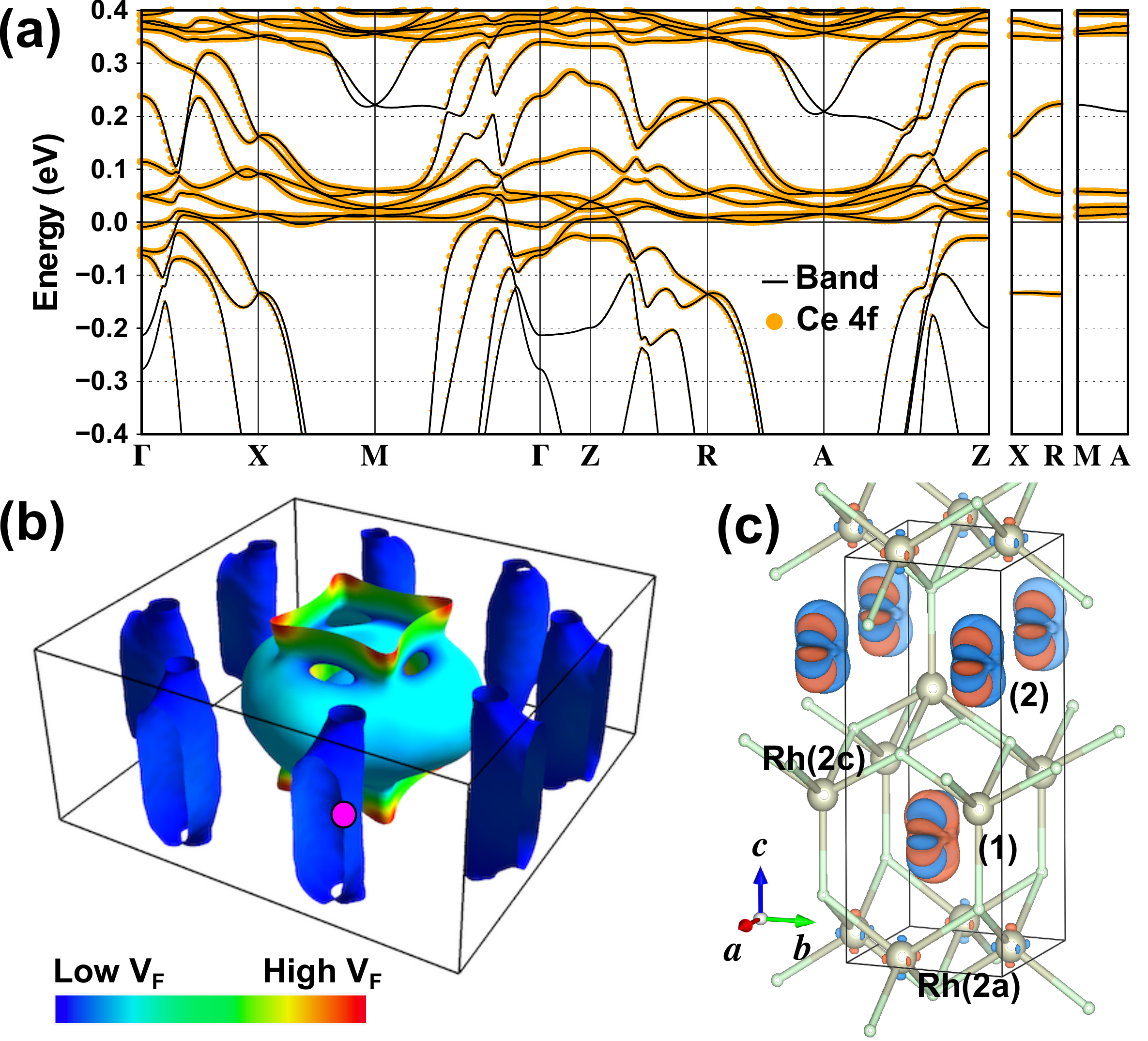}
\caption{(a)
Bands including $4f$ electron correlations through a renormalized band structure
approach with Ce $4f$ weight represented  by orange dots. (b) Fermi surface for (a).
(c) $x$-component spin density distribution of the doubly degenerate
bands at a point on zone edge marked by the pink circle in (b).
Red/blue represents positive/negative spin density.
\label{fig:DFT}
}
\end{figure}

The experimental heat jump at the superconducting transition temperature
suggest
fermion masses a factor 100-1000 larger than the bare electron mass,
implying that the Ce $4f$ electrons are itinerant. The standard DFT results are inconsistent with this enhanced effective mass.
To address this, we have employed a renormalized band structure
approach similar 
to that pioneered by Zwicknagl \cite{Zwicknagl_1992,Zwicknagl_2016}.
Fig.~2a shows the resulting band structure; the corresponding Fermi surface, Fig.~2b,
has a DOS 10 times larger than standard DFT and agrees with that found in
Ref.~\cite{Hafner_2021}. Moreover, the pockets at the zone boundary account for 80\% of
the total DOS, consistent with the observed odd-parity state. In the SM, we show that
with different choices of renormalizations, the DOS can be further increased (with
similar Fermi surfaces), and also explore the effects on the band structure from several other scenarios within DFT. 
Figure \ref{fig:DFT}(c) shows the $S_x$ spin density arising from a Kramers pair on the zone boundary;
the integrated spin
density around each atom is non-vanishing only for $S_x$
and is opposite on the two sublattices, in agreement with the symmetry-based arguments presented above.


{\it Discussion and conclusions---}Our key result is that the
NS P4/nmm structure of CeRh$_2$As$_2$ enables the SOC
structure required to stabilize an odd-parity superconducting state
under field and to enhance the critical field along the $c$-axis.  It is natural to ask if there exist
other materials with the same structure for which this is also the
case. Remarkably, there exist experimental results on superconducting
FeSe, which also crystallizes in a P4/nmm structure, that suggest
similar considerations apply. In particular, Knight shift measurements
indicate that there is no change in the spin susceptibility upon
entering the superconducting state for the field applied along the
$c$-axis \cite{Vinograd_2021,Molatta_2020}. Within the framework we
have discussed here, this could be explained by a nearly-vanishing
g-factor for a $c$-axis Zeeman field due to strong SOC. This implies that the Zeeman coupling only produces a van-Vleck-like spin susceptibility which is largely unchanged by superconductivity \citep{Fischer_2011}. In addition, there exists evidence for an unexplained $c$-axis field-induced superconducting phase transition for fields much larger than the Pauli limiting field \cite{Kasahara_2014}. The possibility that this transition corresponds to  a transition from an even to odd parity phase is currently under investigation.   

{\it Acknowledgements--} MW and TS were supported by the US Department of Energy, Office of Basic Energy Sciences,
Division of Materials Sciences and Engineering under Award
DE-SC0017632. DCC and PMRB were supported by the
Marsden Fund Council from Government funding, managed by Royal Society Te 
Ap\={a}rangi. DFA was supported by the US Department of Energy, Office of Basic Energy Sciences,
Division of Materials Sciences and Engineering under Award
DE-SC0021971. We acknowledge useful discussions with Manuel Brando, Mark Fischer, Christoph Geibel, Elena Hassinger, Seunghyun Khim, Andy Mackenzie, Igor Mazin, and Manfred Sigrist.

\bibliography{Ce122_ref}

\appendix

\section{DFT calculations and analysis}\label{sec:DFT}
DFT calculations, including spin-orbit, for CeRh$_2$As$_2$ are carried out using the Full-Potential Linearized Augmented Plane Wave (FLAPW) 
method  \cite{Mike_FLAPW_2009} with the
structural parameters determined by x-ray diffraction \cite{Ce122_arxiv}.
The Perdew-Burke-Ernzerhof form of the generalized gradient approximation \cite{PBE} is employed for exchange and correlation.
The muffin-tin sphere radii are set to 1.4, 1.2, and 1.1 \AA\  for Ce, Rh, and As atoms, respectively.
The wave function and potential cutoffs are 16 and 200 Ry, respectively.
The Brillouin zone (BZ) is sampled with a 20$\times$20$\times$10 $k$-point mesh during the self-consistent
field cycle.  A denser 50$\times$50$\times$25 mesh is used to determine the Fermi surface, and are
visualized by using {\tt FermiSurfer} \cite{KAWAMURA2019197}.
The crystal structure and spin density distribution are visualized by using {\tt VESTA} \cite{Momma2011}.

\begin{figure}[h]
\includegraphics[width=0.99\columnwidth]{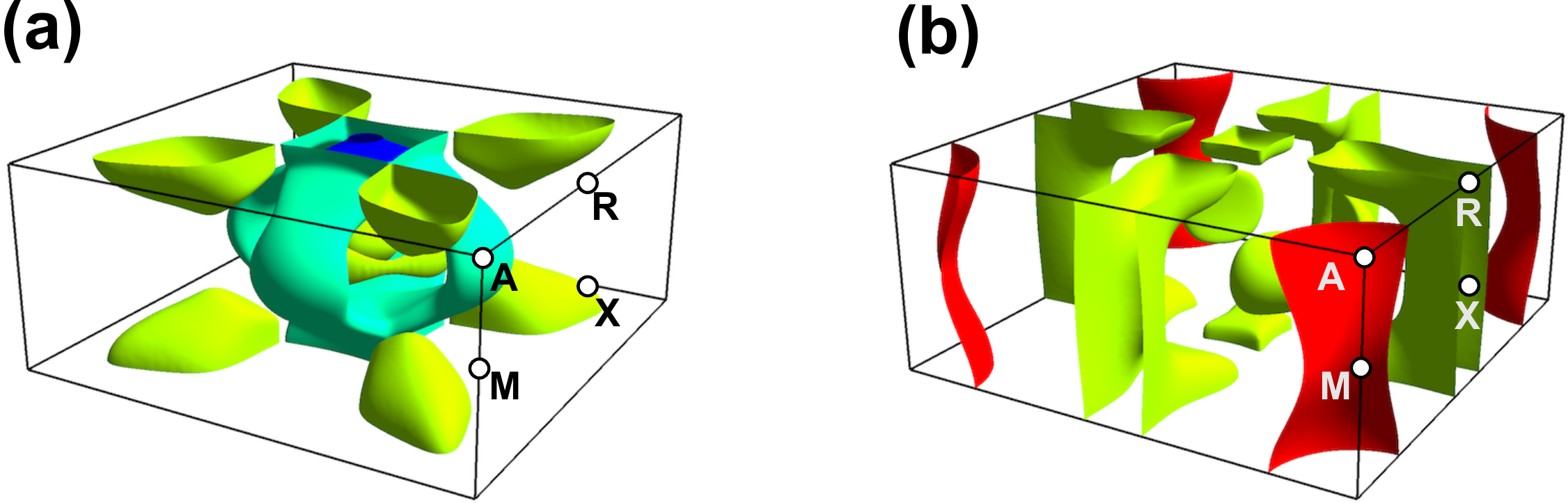}
\caption{\label{fig:supp_FS}
 Fermi surfaces from the standard DFT bands for (a) the calculated Fermi level and (b) for the chemical potential shifted by 0.08 eV. 
}
\end{figure}

The calculated Fermi surface of CeRh$_2$As$_2$ within standard DFT is
shown in Fig.~S\ref{fig:supp_FS}a, and is  in a fair agreement with
previous calculations~\cite{Nogaki_2021}.  
As described in the main text, this Fermi surface does not account for 
the experimental high specific heat jump nor intersects the BZ edge 
where $\tilde{\alpha}_{\bf k}$ diverges. The Fermi surface resulting from a shift of the chemical potential by 0.08 eV, Fig. S1b, has tubes around the M-A line as well as other pieces intersecting the zone edge.

\begin{figure}
\includegraphics[width=0.9\columnwidth]{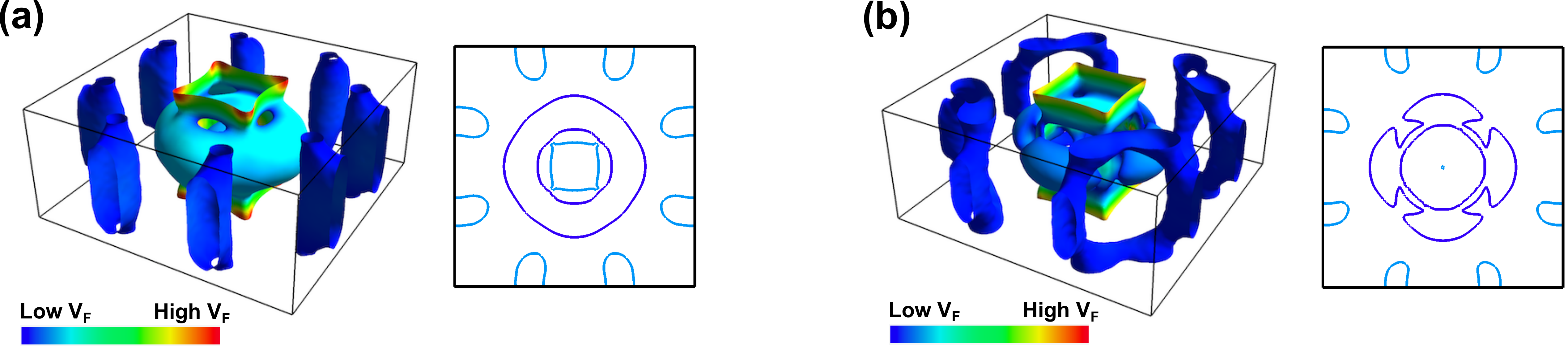}
\caption{\label{fig:supp_RBS}
Fermi surface of CeRh$_2$As$_2$ with ``renormalized'' Ce $4f$ bands. (a)
and (b) correspond to different choices of renormalization paramters,
and provide enhanced DOS at the chemical potential 10 and 20 times
larger, respectively, than the bare DFT. For each case, a 3D view (color
code shows Fermi velocity) and a $k_z$=0 section (two colors stand for two different bands) are shown.
} 
\end{figure}

To partially account for the renormalization/reduction of the bandwidth commonly seen in
4$f$ and heavy fermion materials, we implemented 
a scheme to manipulate
the energy dependence of the logarithmic derivative of the 4$f$
orbitals in the framework of the FLAPW method, an approach closely related to the phase shift technique used in
Ref.~\cite{Hafner_2021}.  Methodological details will be presented elsewhere. 
The resulting calculated 
Fermi surface in Fig.~S\ref{fig:supp_RBS}(a) [Fig. 2(b) of the main
text] closely resembles that in Ref.~\onlinecite{Hafner_2021} with six
tubes attached to the BZ edge.  The total DOS at the Fermi energy is
enhanced by a factor of 10 compared to the bare DFT DOS\@.  
Fig.~S\ref{fig:supp_RBS}(b), obtained from another choice of
renormalization parameters, shows stronger DOS enhancement, now a factor
of 20. The near-edge elements are maintained yet these tubes are now connected at $k_z=\pm\pi$ plane. 
In both cases (and with other choices, not shown here), the Fermi surface elements near the BZ edge constitute 80-90\% of the total DOS with very low Fermi velocity. 

\begin{figure}
\includegraphics[width=\columnwidth]{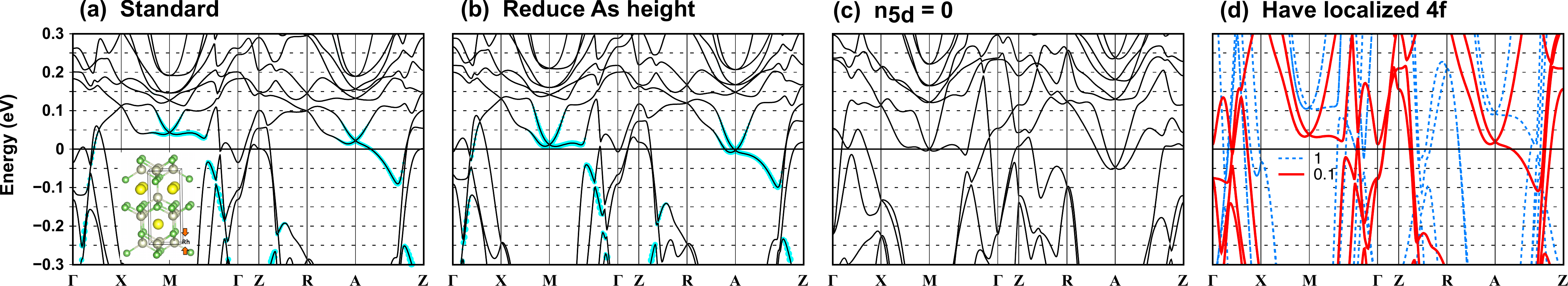}
\caption{\label{fig:supp_bands}
Electronic band structure of CeRh$_2$As$_2$.  (a) Standard calculation using the experimental structural
parameters.  Cyan dots give weight of the $4d_{xy}$ orbital from the $2a$-site Rh (between As atoms).
Inset: an illustrative view of crystal structure where Ce, Rh, and As atoms are shown by yellow, gray, and
green spheres, respectively.  (b) The heights of the As layers sitting above and below the Rh
$2a$ layer are reduced by 0.05\AA\ from the experimental value 1.34\AA.  (c) Positive
infinite potential on the Ce $5d$ orbitals to illustrate $5d$-$4f$ valence fluctuation. The $5d$ electrons
(0.70 electrons within a Ce-nucleus centered sphere of radius 1.4\AA) are eliminated from the occupied
levels while the $4f$ count is increased by 0.54.  (d) Localized (core) $4f$ states for 1 or 0.1 $4f$
electrons to illustrate the effects of localization and $f^1 \to f^0$ mixed valence.  
}
\end{figure}

In addition, we examine other possible scenarios, within standard DFT, that
alter the relative energies of the bands and provide Fermi surface elements near the BZ edge. 
The standard DFT calculation shows that the Ce $4f$ states predominantly hybridize with the conduction Rh $4d$ bands.  
In particular, the bottom of the conduction band around the M-A Dirac line of the Brillouin zone is of
particular interest since this band, having Ce $4f$ contributions, shows large Rashba-type band splitting.  
This state is built up of Rh $xy$ (and $z^2$) orbitals on the $2a$ site
[the site sandwiched by As atoms; see inset of Fig.~S\ref{fig:supp_bands}(a)] antibonded to As $p$ states.   
Therefore, the energy position of this band (which at $M$ point is located about 0.05 eV above the Fermi
level) is sensitive to the As-layer height relative to the Rh $2a$ layer; as demonstrated in 
Figure~S\ref{fig:supp_bands}(b), when the As height is slightly reduced (by 0.05 \AA), this band moves
down to the Fermi level.  The energy position of this $4f$-hybridized band is also sensitive to the mixed
valent and  localized/delocalized nature of the $4f$ orbital \cite{varma76,npjqm2018}.  
Figure~S\ref{fig:supp_bands}(c) shows the effect of Ce $5d$-$4f$ ($f^1\to f^2$) valence fluctuation, which
is simulated by adding infinite potential on the $5d$ state. To mimic localized $4f$ orbitals and $f^1\to
f^0$ fluctuations, the 4$f$
states are 
treated as core electrons (with the valence states properly orthogonalized), and are shown in
Fig.~S\ref{fig:supp_bands}(d). In all of these cases, the bands around M-A near the chemical potential are
common features, although the exact position relative to the chemical potential does vary.
Importantly, the robustness of these features implies that the essential physics of the staggered Rashba
splitting about the M-A Dirac line near the chemical potential remains unchanged.

\section{Pseudospin basis}\label{sec:pseudo}

The defining feature of the pseudospin index is that it transforms
like a spin under inversion and time-reversal symmetry. That is,
letting $\{|{\bm k},b,\Uparrow \rangle,\, |{\bm k},b,\Downarrow
\rangle\}$ be the pseudospin basis at momentum ${\bm k}$ in band $b$,
we have the operation of inversion
\begin{equation}
P|{\bm k},b,\Uparrow \rangle = |-{\bm k},b,\Uparrow
\rangle\,, \quad P|{\bm k},b,\Downarrow \rangle = |-{\bm k},b,\Downarrow
\rangle\,,
\end{equation}
 and time-reversal 
\begin{equation}
T|{\bm k},b,\Uparrow
\rangle = -|-{\bm k},b,\Downarrow \rangle\,, \quad T|{\bm k},b,\Downarrow
\rangle = |-{\bm k},b,\Uparrow \rangle\,.
\end{equation}
It is furthermore often convenient to adopt a pseudospin basis where the
pseudospin has the same transformation properties as the electronic
spin under the point group operations, a so-called manifestly
covariant Bloch basis (MCBB)~\cite{Fu_MCBB}.
Consider a symmetry operation $g$ of the point group
such that
\begin{equation}
U_gH_0({\bf k})U_g^\dagger = H_0(g{\bf k}) \,,
\end{equation}
where $U_g$ is the unitary matrix for the symmetry
operation in the four-component basis. The eigenvectors
$\phi_{{\bf k},\pm,s}$ define an MCBB if the matrix with
columns composed of these vectors,
\begin{equation}
\Psi_{\bf k} = (\psi_{{\bf k},+,\Uparrow},
  \psi_{{\bf k},+,\Downarrow},
  \psi_{{\bf k},-,\Uparrow},
  \psi_{{\bf k},-,\Downarrow}) \,,
\end{equation}
satisfies
\begin{equation}
\Psi_{g{\bf k}}^\dagger U_{g}\Psi_{{\bf k}}
  = {s}_0\otimes u_g \,,
\end{equation}
where $u_{g}$ is the equivalent symmetry operation for
a spin-$1/2$ system.

We adopt the following pseudospin basis for our model
\begin{widetext}
\begin{eqnarray}
\psi_{{\bm k},+,\Uparrow} = 
\left(\begin{array}{c}
-e^{-\frac{i \xi_{\bm k} }{2}} \cos \left(\frac{\chi_{\bm k} }{2}\right) \cos
        \left(\frac{\omega_{\bm k} }{2}\right) \\  -e^{-\frac{1}{2} i (\xi_{\bm k} -2
                                         \phi_{\bm k} )} \sin \left(\frac{\chi_{\bm k}
                                         }{2}\right) \cos
                                         \left(\frac{\omega_{\bm k}
                                         }{2}\right) \\ -e^{\frac{i \xi_{\bm k} }{2}} \cos \left(\frac{\chi_{\bm k} }{2}\right) \sin \left(\frac{\omega_{\bm k}
   }{2}\right) \\  e^{\frac{1}{2} i (\xi_{\bm k} +2 \phi_{\bm k} )} \sin \left(\frac{\chi_{\bm k}
        }{2}\right) \sin \left(\frac{\omega_{\bm k} }{2}\right)
\end{array}\right) &\qquad & \psi_{{\bm k},+,\Downarrow} =
                            \left(\begin{array}{c}
-e^{-\frac{1}{2} i (\xi_{\bm k} +2 \phi_{\bm k} )} \sin \left(\frac{\chi_{\bm k} }{2}\right)
                                    \sin \left(\frac{\omega_{\bm k}
                                    }{2}\right) \\ -e^{-\frac{i \xi_{\bm k}
                                    }{2}} \cos \left(\frac{\chi_{\bm k}
                                    }{2}\right) \sin
                                    \left(\frac{\omega_{\bm k} }{2}\right) \\ e^{\frac{1}{2} i (\xi_{\bm k} -2 \phi_{\bm k} )} \sin \left(\frac{\chi_{\bm k} }{2}\right) \cos \left(\frac{\omega_{\bm k}
   }{2}\right)\\ -e^{\frac{i \xi_{\bm k} }{2}} \cos \left(\frac{\chi_{\bm k}
                                    }{2}\right) \cos
                                    \left(\frac{\omega_{\bm k}
                                    }{2}\right)\end{array}\right)\\
\psi_{{\bm k},-,\Uparrow} =  \left(\begin{array}{c}
-ie^{-\frac{i \xi_{\bm k} }{2}} \text{sgn}(\xi_{\bm k} ) \cos \left(\frac{\chi_{\bm k}
                                        }{2}\right) \sin
                                        \left(\frac{\omega_{\bm k}
                                        }{2}\right)\\  ie^{-\frac{1}{2}
                                                     i (\xi_{\bm k} -2 \phi_{\bm k} )}
                                                     \text{sgn}(\xi_{\bm k} )
                                                     \sin
                                                     \left(\frac{\chi_{\bm k}
                                                     }{2}\right) \sin
                                                     \left(\frac{\omega_{\bm k}
                                                     }{2}\right) \\  ie^{\frac{i \xi_{\bm k} }{2}} \text{sgn}(\xi_{\bm k} ) \cos \left(\frac{\chi_{\bm k}
   }{2}\right) \cos \left(\frac{\omega_{\bm k} }{2}\right) \\  ie^{\frac{1}{2}
                                        i (\xi_{\bm k} +2 \phi_{\bm k} )}
                                        \text{sgn}(\xi_{\bm k} ) \sin
                                        \left(\frac{\chi_{\bm k} }{2}\right)
                                        \cos \left(\frac{\omega_{\bm k}
                                        }{2}\right)\end{array} \right)&&
\psi_{{\bm k},-,\Downarrow} =  \left(\begin{array}{c}
 ie^{-\frac{1}{2} i (\xi_{\bm k} +2 \phi_{\bm k} )} \text{sgn}(\xi_{\bm k} ) \sin
                                        \left(\frac{\chi_{\bm k} }{2}\right)
                                        \cos \left(\frac{\omega_{\bm k}
                                        }{2}\right)\\ - ie^{-\frac{i
                                        \xi_{\bm k} }{2}} \text{sgn}(\xi_{\bm k} )
                                        \cos \left(\frac{\chi_{\bm k}
                                        }{2}\right) \cos
                                        \left(\frac{\omega_{\bm k}
                                        }{2}\right)\\  ie^{\frac{1}{2} i (\xi_{\bm k} -2 \phi_{\bm k} )} \text{sgn}(\xi_{\bm k} ) \sin
   \left(\frac{\chi_{\bm k} }{2}\right) \sin \left(\frac{\omega_{\bm k} }{2}\right) \\
                                         ie^{\frac{i \xi_{\bm k} }{2}}
                                         \text{sgn}(\xi_{\bm k} ) \cos
                                         \left(\frac{\chi_{\bm k}
                                         }{2}\right) \sin
                                         \left(\frac{\omega_{\bm k}
                                         }{2}\right)\end{array}
  \right) \notag \\
\end{eqnarray}
\end{widetext}
where the angles are defined in terms of the coefficients of the
general Hamiltonian as
\begin{align}
  \label{eq:angles2}
\xi_{\bm k} & = \arctan\left(\frac{\epsilon_{y0,{\bf
              k}}}{\epsilon_{x0,{\bm k}}}\right) \\
\omega_{\bm k} & = \arctan\left(\frac{\sqrt{\epsilon_{x0,{\bm k}}^2 +
    \epsilon_{y0,{\bm k}}^2}}{\epsilon_{zz,{\bm k}}}\right) \\
\chi_{\bm k} & = \arctan\left(\frac{\sqrt{\epsilon_{zx,{\bf
        k}}^2 + \epsilon_{zy,{\bm k}}^2}}{\sqrt{\epsilon_{x0,{\bm k}}^2 +
    \epsilon_{y0,{\bm k}}^2 + \epsilon_{zz,{\bm k}}^2}}\right)\\
  \phi_{\bm k} & = \arctan\left(\frac{\epsilon_{zy,{\bf
                 k}}}{\epsilon_{zx,{\bm k}}}\right)
\end{align}
Note that the appearance of $\text{sgn}(\xi_{\bm k} )$ in the
definition of the $-$ band states is ill-defined if $\xi_{\bm k}=0$;
since $\xi_{\bm k} = -\xi_{-{\bm k}}$, however, on this set of
  measure zero we can
define $\text{sgn}(\xi_{\bm k})=1$ and $\text{sgn}(\xi_{-\bf
  k})=-1$. 

A key feature of this pseudospin basis is that it converges smoothly
to the electronic spin in the limit of vanishing SOC, which
is reached when $\omega_{\bm k} \rightarrow \frac{\pi}{2}$ and
$\chi_{\bm k} \rightarrow 0$. Other important cases are a purely Rashba
SOC ($\epsilon_{zz,{\bm k}}=0$), where  we
have $\omega_{\bm k}=\frac{\pi}{2}$, and a purely Ising
SOC ($\epsilon_{zx,{\bm k}}=\epsilon_{zy,{\bm k}}=0$)
which is characterized by $\chi_{\bm k}=0$. In the latter case the
pseudospin is again equivalent to the electronic spin.

We now consider the projection of the spin operators onto the
pseudospin basis of each band:
\begin{eqnarray}
\tau_0\sigma_x & \rightarrow &
\sin(\omega_{\bm k})\left(\cos(\chi_{\bm k})\sin^2(\phi_{\bm k})  +
\cos^2(\phi_{\bm k})\right)s_x \\ && 
+  \sin^2\left(\frac{\chi_{\bm k} }{2}\right) \sin (\omega_{\bm k} ) \sin (2 \phi_{\bm k}) s_y   
\\&& +  \sin (\chi_{\bm k} ) \cos (\omega_{\bm k} ) \cos (\phi_{\bm k} )s_z\\
\tau_0\sigma_y & \rightarrow & \sin^2\left(\frac{\chi_{\bm k}}{2}\right)\sin (\omega_{\bm k} ) \sin(2\phi_{\bm k} )s_x \\ &&  
 +\sin(\omega_{\bm k})\left(\sin^2(\phi_{\bm k}) +\cos(\chi_{\bm k})\cos^2(\phi_{\bm k})\right)s_y 
\\&& + \sin (\chi_{\bm k} ) \cos (\omega_{\bm k} ) \sin (\phi_{\bm k} )s_z\\
\tau_0\sigma_z & \rightarrow & \cos(\chi_{\bm k})s_z 
\end{eqnarray}

The pseudospin and Zeeman fields are not generally
colinear, with the notable exception of a $c$-axis Zeeman field. In
both the Rashba and Ising limits in-plane and out-of-plane
Zeeman fields produce pseudospin fields which are also strictly in-plane and
out-of-plane, respectively.

We now turn to the staggered singlet pairing state. Expressed in the
pseudospin basis we have 
\begin{align}
  \tau_zi\sigma_y \rightarrow  &\pm\left[\sin\chi_{\bm k}\sin\omega_{\bm k}\left(\cos\phi_{\bm k}\hat{s}_x +
\sin\phi_{\bm k}\hat{s}_y\right) + \cos\omega_{\bm k} \hat{s}_z\right]i\hat{s}_y 
\end{align}
Since we have a MCBB, the effective ${\bf d}$-vector indeed transforms
as a state in the $A_{2u}$ irrep, with ${\bf d}_{\bm k} \sim a_1 (f_y
\hat{\bf x} - f_x\hat{\bf y}) + a_2 f_{A_{1u}}{\bf z}$, where
$f_\nu$ has the transformation properties of a $p_\nu$-harmonic,
whereas $f_{A_{1u}}$ transforms as $A_{1u}$. We note that in the Ising
limit the effective ${\bf d}$-vector is oriented along the $c$-axis,
whereas in the Rashba limit the ${\bf d}$-vector is purely
in-plane. Comparing this to the projected spin operators, we note the
remarkable result that the staggered singlet state is immune to
in-plane and out-of-plane fields in the Ising and Rashba limits,
respectively. This is a particular feature of the staggered
singlet state which makes it generically robust against applied
fields. 

\section{Model calculation}\label{sec:model}


\begin{figure}
\centering
\begin{overpic}[width=0.4\columnwidth]{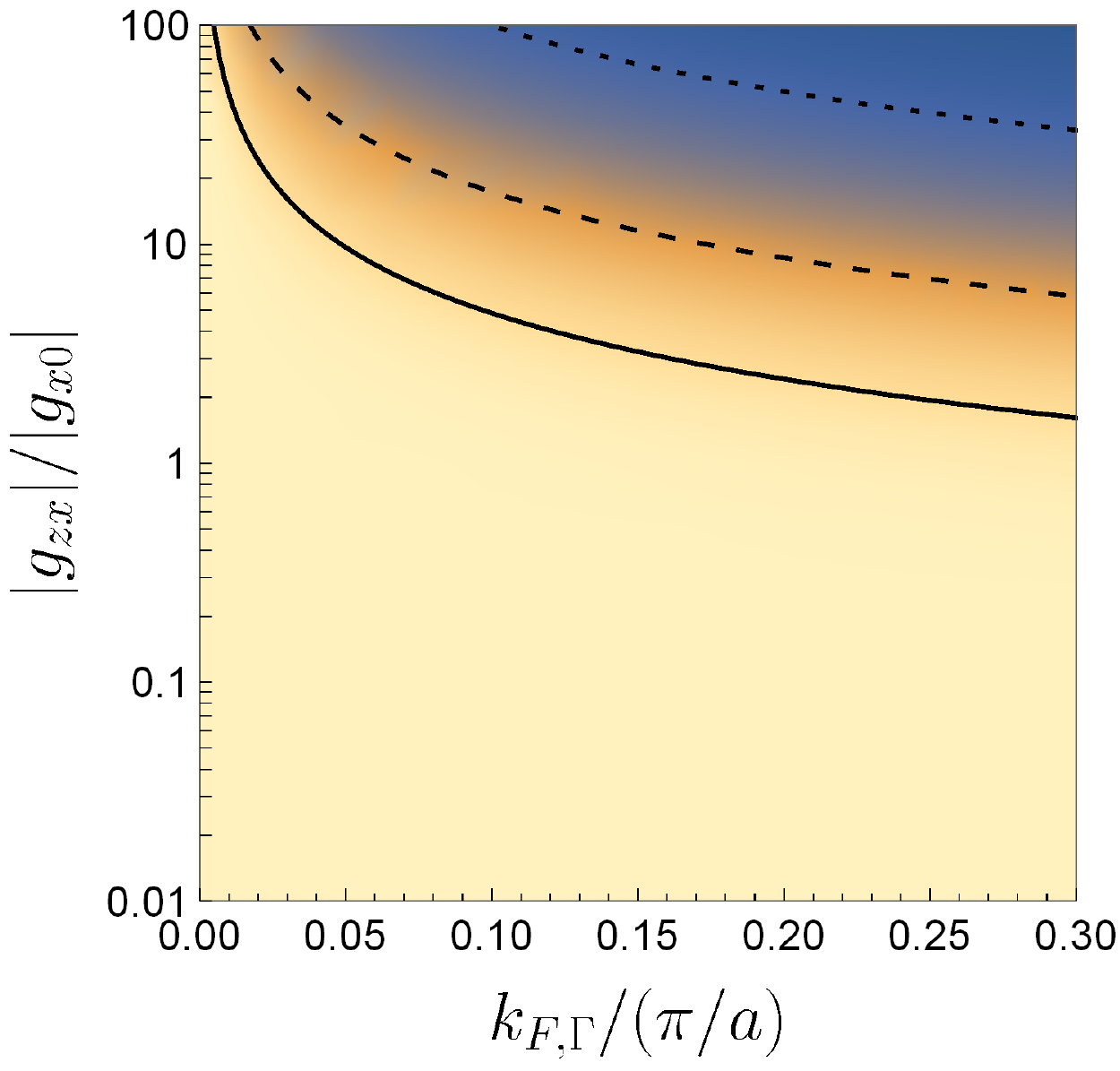}
\put(0,90){(a)}
\end{overpic}
\begin{overpic}[width=0.4\columnwidth]{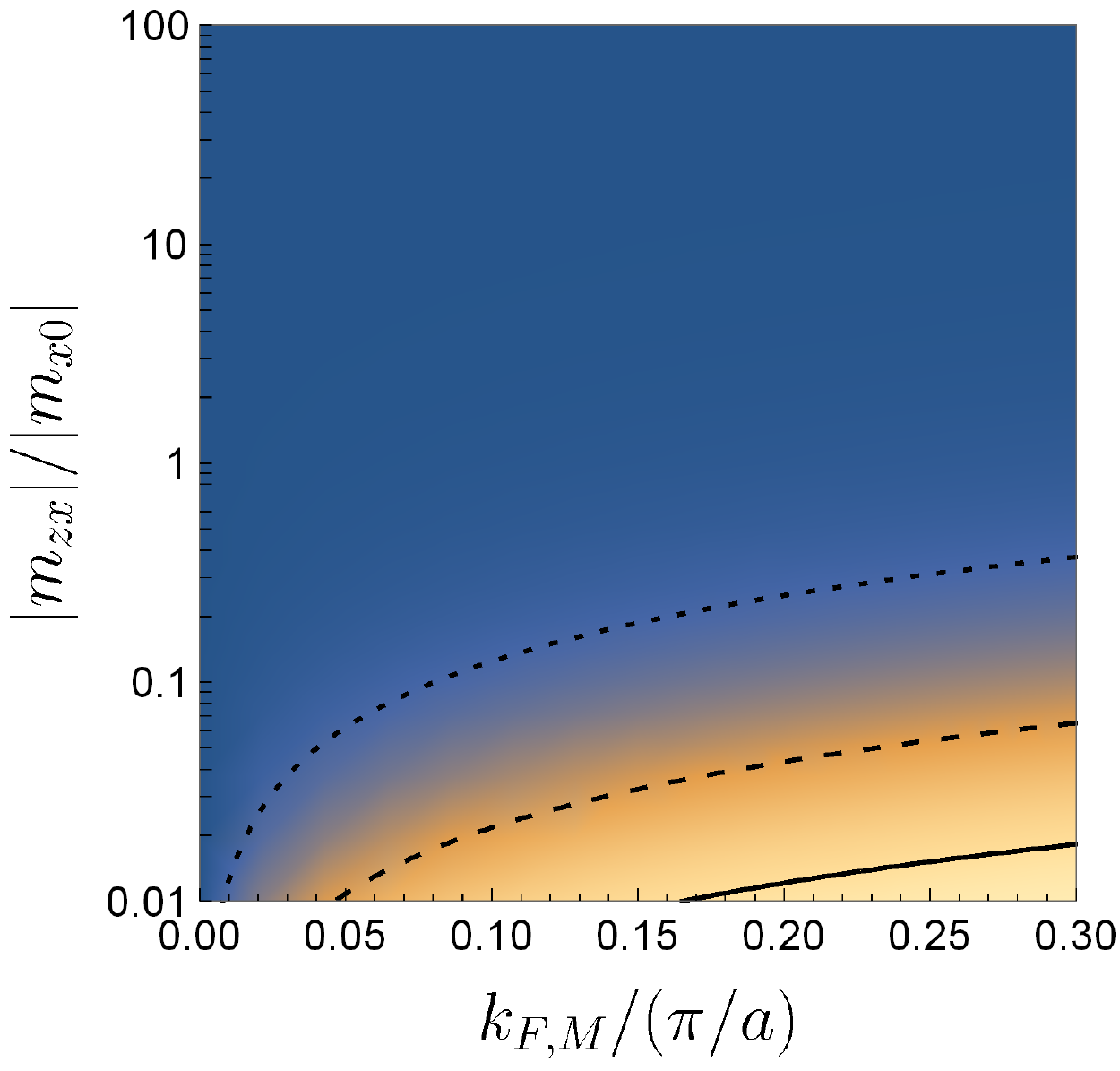}
\put(0,90){(b)}
\end{overpic}\\
\begin{overpic}[width=0.35\textwidth]{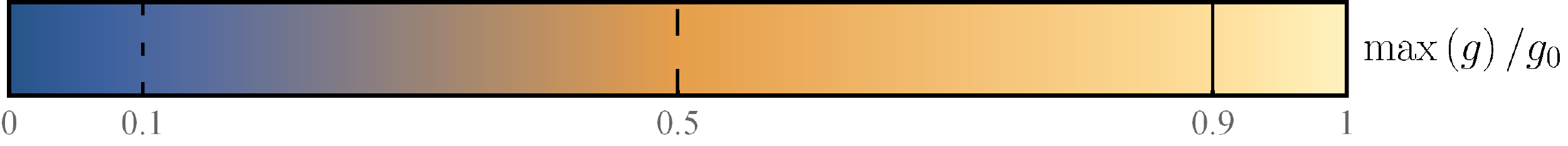}
\end{overpic}
\caption{ The dependence of the effective $g$-factor magnitude on the $\Gamma$ point Fermi surface pocket (a) and maximum magnitude on the $M$ point Fermi surface pocket (b) on the ratio of SOC to interlayer hopping and relevant Fermi wavevector $k_F$. }\label{fig:appendix_g_factors}
\end{figure}

In a $c$-axis field $H$, the free energy of the uniform singlet superconductor relative to the
normal state is given by
\begin{equation}
F_S = \mathcal{N}_0\left|\Delta_0\right|^2\Phi\left(\frac{\Delta}{\Delta_0},\frac{k_B T}{\Delta_0},\frac{g\mu_B H}{\Delta_0}\right),\label{eq:app_FreeEnergy}
\end{equation}
where $\mathcal{N}_0$ is the total density of state at the Fermi
energy, $\Delta_0$ is the zero temperature and zero field
superconducting gap. The latter is explicitly given by  
\begin{widetext}
\begin{equation}
\Delta_0 = 2\Lambda\exp\left(-\frac{1}{V_0\sum_{\nu}\mathcal{N}_{\nu}\left\langle \left|f_{\nu,\bm{k}}\right|^2\right\rangle_{\nu}}\right)\exp\left(-\frac{\sum_{\nu}\mathcal{N}_{\nu}\left\langle \left|f_{\nu,\bm{k}}\right|^2\log\left(\left|f_{\nu,\bm{k}}\right|\right)\right\rangle_{\nu}}{\sum_{\nu}\mathcal{N}_{\nu}\left\langle \left|f_{\nu,\bm{k}}\right|^2\right\rangle_{\nu}}\right),\label{eq:app_Delta0}
\end{equation}
where $\Lambda$ is the energy scale for the pairing interaction (the
exact magnitude of which does not enter into our calculations), $V_0$
is the pairing interaction. The $\nu$ index in
Eqn. \ref{eq:app_Delta0} runs over the Fermi surfaces, and $\langle\ldots\rangle_{\nu}$ denotes the average over the $\nu$ Fermi surface,  with $\mathcal{N}_{\nu}$ the density of states and $f_{\nu,\bm{k}}$ the form factor of the superconducting gap in Fermi surface $\nu$. In Eqn. \ref{eq:app_FreeEnergy}, we have also defined the dimensionless function
\begin{equation}
\Phi\left(\delta,t,h\right) = \sum_{\nu}\frac{\mathcal{N}_{\nu}}{\mathcal{N}_0} \int\limits_{0}^{\infty}dx \left\langle \frac{\left|f_{\nu,\bm{k}}\right|^2\delta^2}{\sqrt{x^2+\left|f_{\nu,\bm{k}}\right|^2}} - 2t\sum_{\sigma=\pm}\log\left[\frac{\cosh\left(\frac{\sigma\left|\vec{\gamma}_{\nu,\bm{k}}^z\right| h +\sqrt{x^2+\left|f_{\nu,\bm{k}}\right|^2\delta^2}}{2t}\right)}{\cosh\left(\frac{\sigma\left|\vec{\gamma}_{\nu,\bm{k}}^z\right| h +x}{2t}\right)}\right] \right\rangle_{\nu} \label{eq:app_Phi}
\end{equation}
with  $\left|\vec{\gamma}_{\nu,\bm{k}}^z\right|$ the
(momentum-dependent) effective $g$-factor on the $\nu$ Fermi surface
pocket. For the pocket centered about the M-A Dirac line, the effective
$g$-factor vanishes with the radius of the pocket
$k_{F,MA}\rightarrow 0$, while for the pocket at the $\Gamma$-Z line the effective
$g$-factor converges to the free electron value as
$k_{F,\Gamma{Z}}\rightarrow 0$ due to the vanishing of the ratio
$\tilde{\alpha}_{\bm k}$. 
Additionally, the effective
$g$-factor of the $M$ point Fermi surface pocket is strongly
anisotropic. From our $\kp$ theory at $k_z=0$, the dependence of the effective $g$-factors on the model
parameters and Fermi pocket size are presented in
Fig.~S\ref{fig:appendix_g_factors}. Clearly, for a wide parameter range the
maximum of the $g$-factor for the M-A pocket is typically close to
zero, while for the $\Gamma$-Z pocket the $g$-factor is generally only
slightly smaller than $g_0$. 

The free energy of the staggered singlet (relative to the normal state) is similarly given by
\begin{equation}
F_{S,s} = \mathcal{N}_0\left|\Delta_{0,s}\right|^2\Phi_{s}\left(\frac{\Delta}{\Delta_{0,s}},\frac{k_B T}{\Delta_{0,s}}\right),\label{eq:app_stagFE}
\end{equation}
where $\Phi_s$ is related to Eq. \ref{eq:app_Phi} by
$\Phi_{s}(\delta,t) = \Phi(\delta,t,h=0)$ with the substitution
$\left|f_{\nu,\bm{k}}\right| \rightarrow
\sqrt{r_{\nu,\bm{k}}}\left|f_{\nu,\bm{k}}\right|$ which accounts for
the reduced magnitude of the gap on the Fermi surface, with
$r_{\nu,\bm{k}}$ the superconducting fitness on band $\nu$
\cite{Ramires2018}. The $c$-axis magnetic field does not
enter this expression since it is not pair breaking for the staggered
singlet state,  Assuming the pairing interaction $V_0$ is of equal
strength in both the uniform and staggered singlet channels, the
pairing potential at zero temperature is
\begin{equation}
\Delta_{0,s} = 2\Lambda\exp\left(-\frac{1}{V_0\sum_{\nu}\mathcal{N}_{\nu}\left\langle r_{\nu,\bm{k}}\left|f_{\nu,\bm{k}}\right|^2\right\rangle_{\nu}}\right)\exp\left(-\frac{\sum_{\nu}\mathcal{N}_{\nu}\left\langle r_{\nu,\bm{k}} \left|f_{\nu,\bm{k}}\right|^2\log\left(\sqrt{r_{\nu,\bm{k}}} \left|f_{\nu,\bm{k}}\right|\right)\right\rangle_{\nu}}{\sum_{\nu}\mathcal{N}_{\nu}\left\langle r_{\nu,\bm{k}}\left|f_{\nu,\bm{k}}\right|^2\right\rangle_{\nu}}\right).
\end{equation}
The necessary choice of a specific value for the pairing interaction is the only non-universal aspect of our calculation, as the energy scale $\Lambda$ is irrelevant since the staggered singlet free energy can be expressed in terms of the ratio between the two gap magnitudes, and we assume an intermediate pairing interaction strength of $V_0\mathcal{N}_0=0.3$. 
\end{widetext}

\begin{figure}
\begin{overpic}[width=0.35\columnwidth]{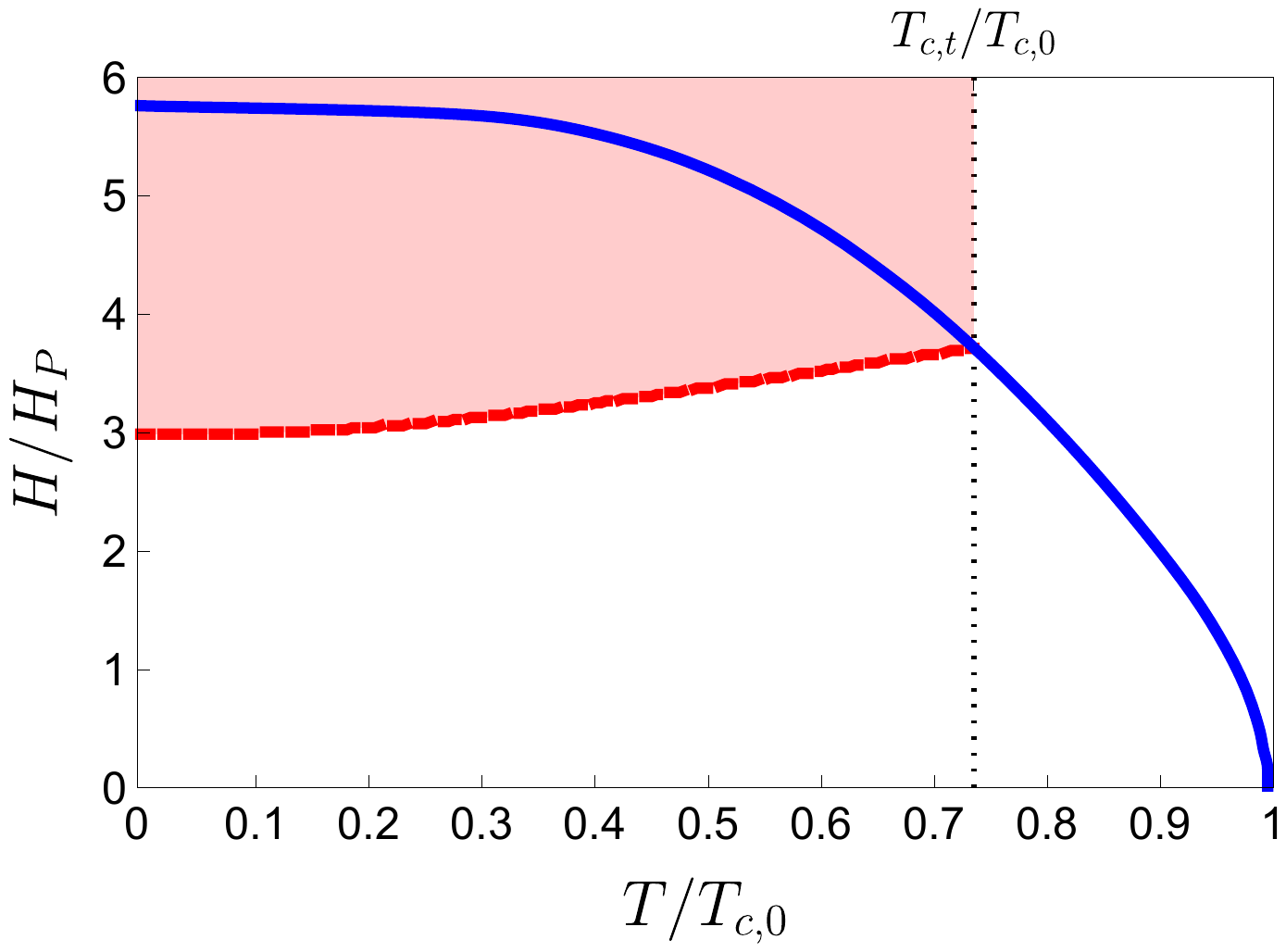}
\put(0,0){(a)}
\end{overpic}
\begin{overpic}[width=0.35\columnwidth]{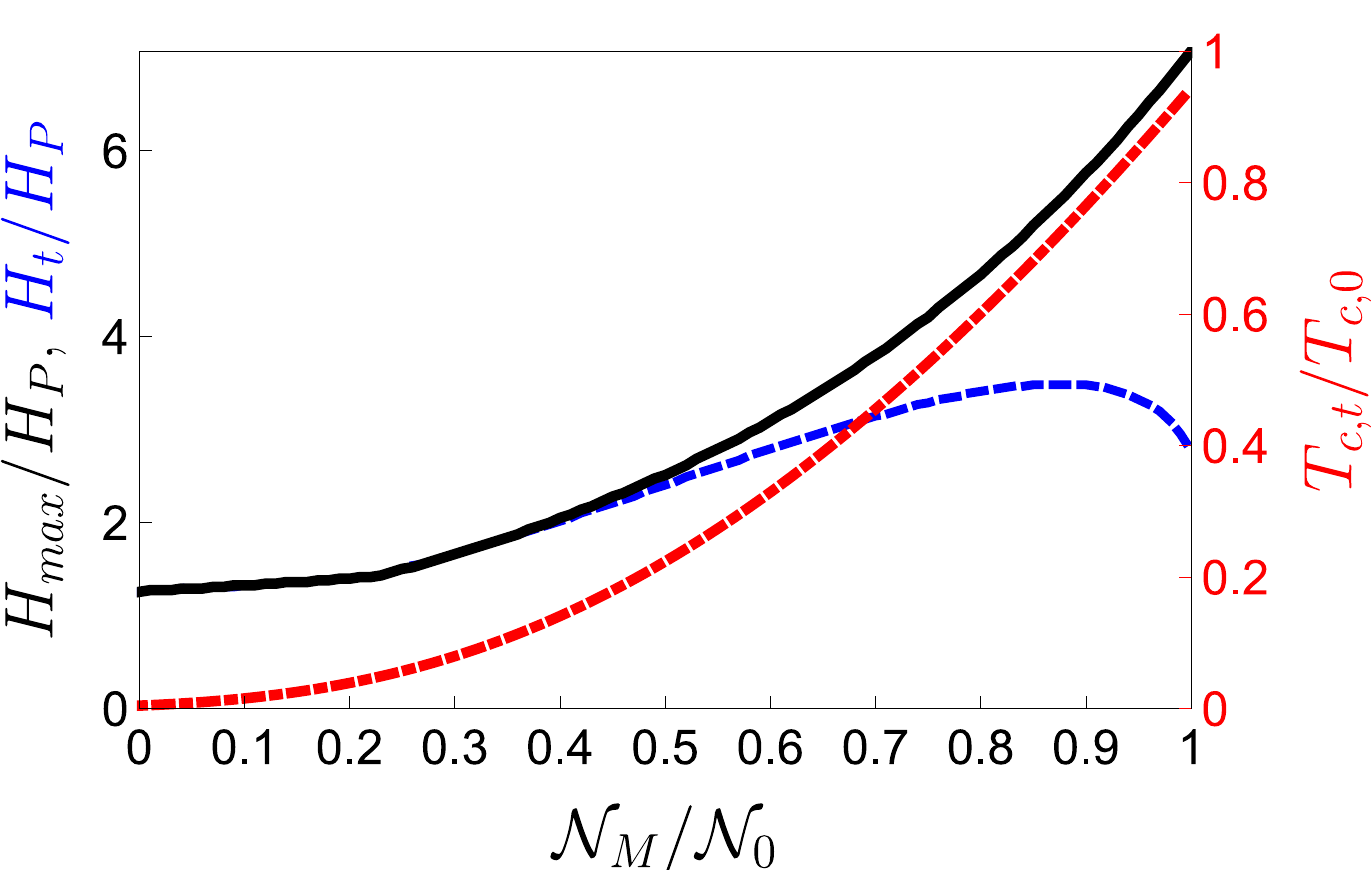}
\put(0,0){(b)}
\end{overpic}
\caption{Variation of the effective $g$-factor magnitudes: (a) The phase diagram
for $\mathcal{N}_M=0.9\mathcal{N}_0$, the blue line gives the upper critical field
of the uniform state, and the red dashed line gives the boundary of
the staggered state (shaded). (b) Plot showing the dependence of the upper critical field of
the uniform state ($H_{\text{max}}$) and the critical field ($H_t$)
and temperature ($T_{c,t}$) at which the field-induced transition
occurs as a function of the contribution of the density of states at
the $M$ point to the total density of states, assuming an equal pairing interaction strength in both channels, and $g_{\Gamma}=0.8 g_0$ and $\text{max}(g_{M})=0.2g_0$.}\label{fig:appendix_different_g_factors}
\end{figure}

We minimize the free energy of the uniform and staggered singlet
states, Eqs. \ref{eq:app_FreeEnergy} and \ref{eq:app_stagFE}, and
compare the two over the range of temperature and field to obtain the
phase diagram. 
Reasonable variation of the effective $g$-factor magnitudes  or
  interaction strength does not qualitatively alter the phase
diagram.  However, the reduction  of the effective $g$-factor at the $\Gamma$ point and enhancement at the $M$ point in Fig.~S \ref{fig:appendix_different_g_factors} reduces the critical field magnitudes, but does not qualitatively alter the phase diagram otherwise. 

\begin{figure}
\begin{overpic}[width=0.35\columnwidth]{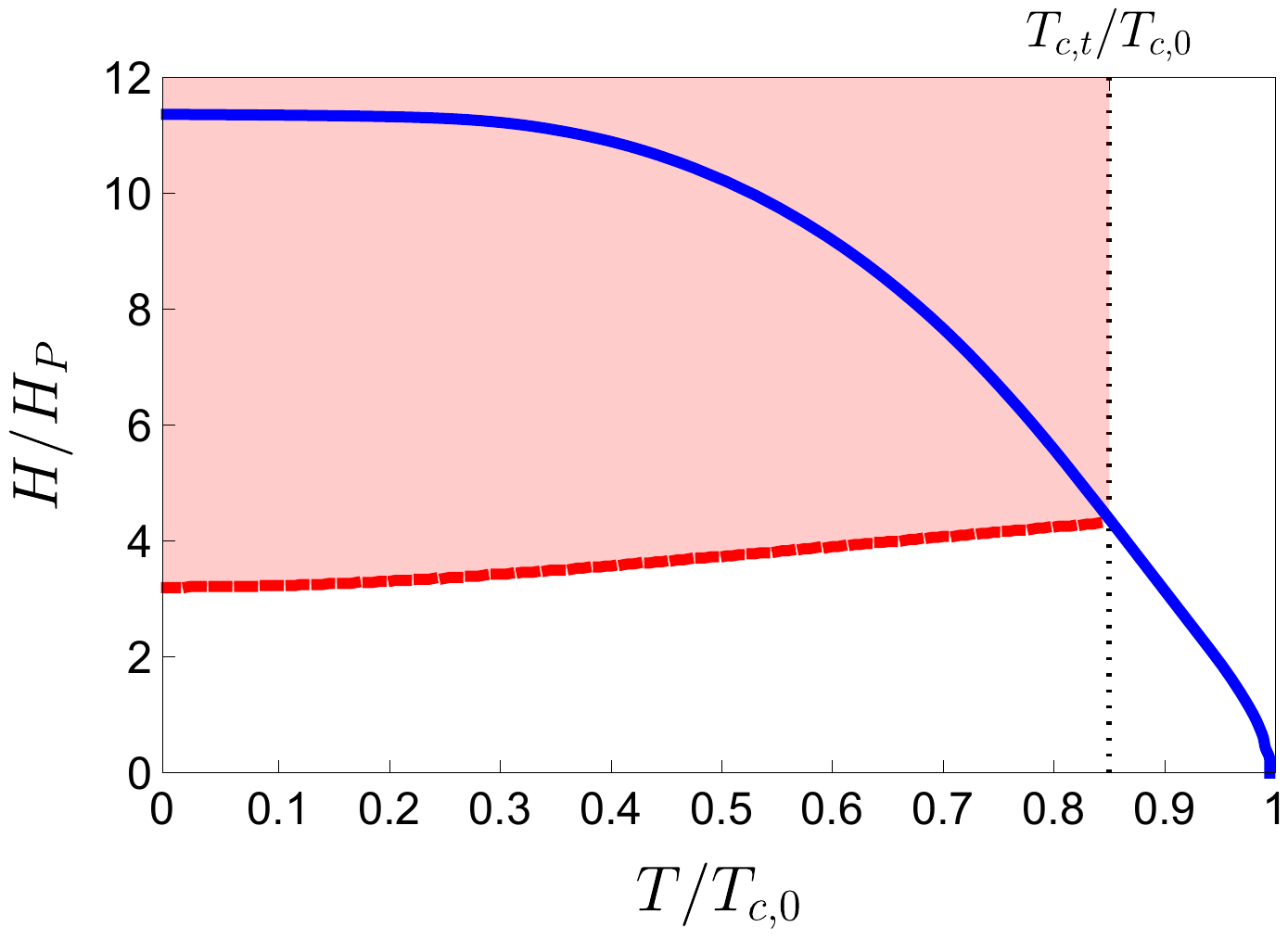}
\put(0,0){(a)}
\end{overpic}
\begin{overpic}[width=0.35\columnwidth]{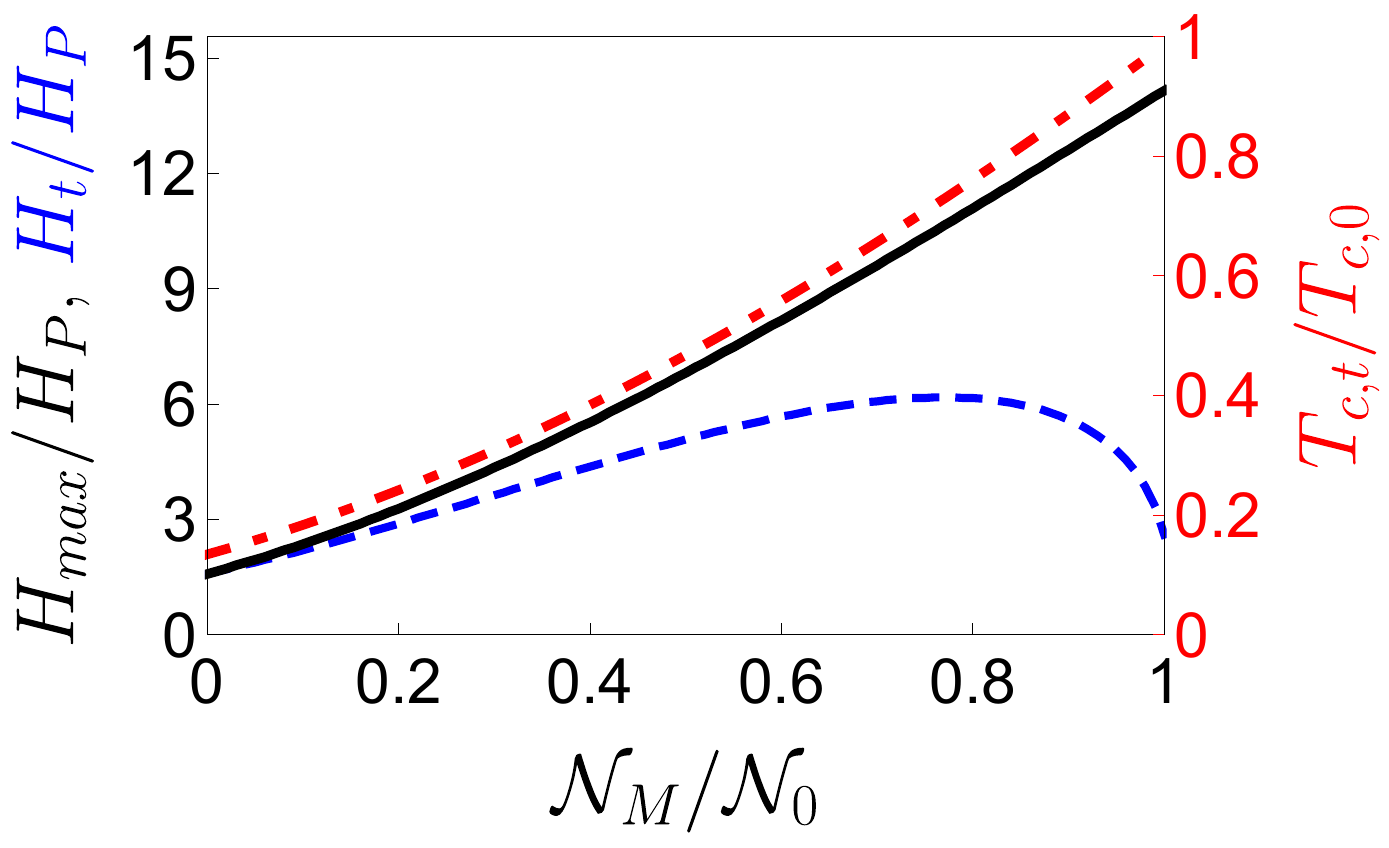}
\put(0,0){(b)}
\end{overpic}
\caption{Two pocket model with pockets at $M$-$A$ and $X$-$R$: (a) The phase diagram
for $\mathcal{N}_M=0.9\mathcal{N}_0$, the blue line gives the upper critical field
of the uniform state, and the red dashed line gives the boundary of
the staggered state (shaded). (b) Plot showing the dependence of the upper critical field of
the uniform state ($H_{\text{max}}$) and the critical field ($H_t$)
and temperature ($T_{c,t}$) at which the field-induced transition
occurs as a function of the contribution of the density of states at
the $M$ point to the total density of states, assuming an equal pairing interaction strength in both channels, and $g_{X}=0.9 g_0$ and $\text{max}(g_{M})=0.1g_0$.  }\label{fig:appendix_X-M}
\end{figure}

We can additionally consider a model with Fermi surface pockets around
the M-A and X-R Dirac  lines, where the latter pocket has a strongly
anisotropic effective $g$-factor which is independent of
$k_{F,\text{XR}}$. For reasonable ratios of the SOC
strength 
and interorbital hopping integral it has a maximum value close to the
bare $g_0$, although the anisotropy significantly reduces the average
value of the effective $g$-factor. In this model, we find that the M-A
pocket is still the most significant contributor to the stability of
the staggered singlet state, but that the pocket about X-R is
considerably less detrimental to the staggered singlet state than the
$\Gamma$-Z pocket, due to the smaller average $g$-factor in the
former. In Fig.~S\ref{fig:appendix_X-M} we highlight the effect of
including an X-R pocket instead of the $\Gamma$-Z pocket, and see that
the transition to the staggered singlet state generically occurs at a
higher value of $T_{c,t}$. 

\begin{figure}
\begin{overpic}[width=0.35\columnwidth]{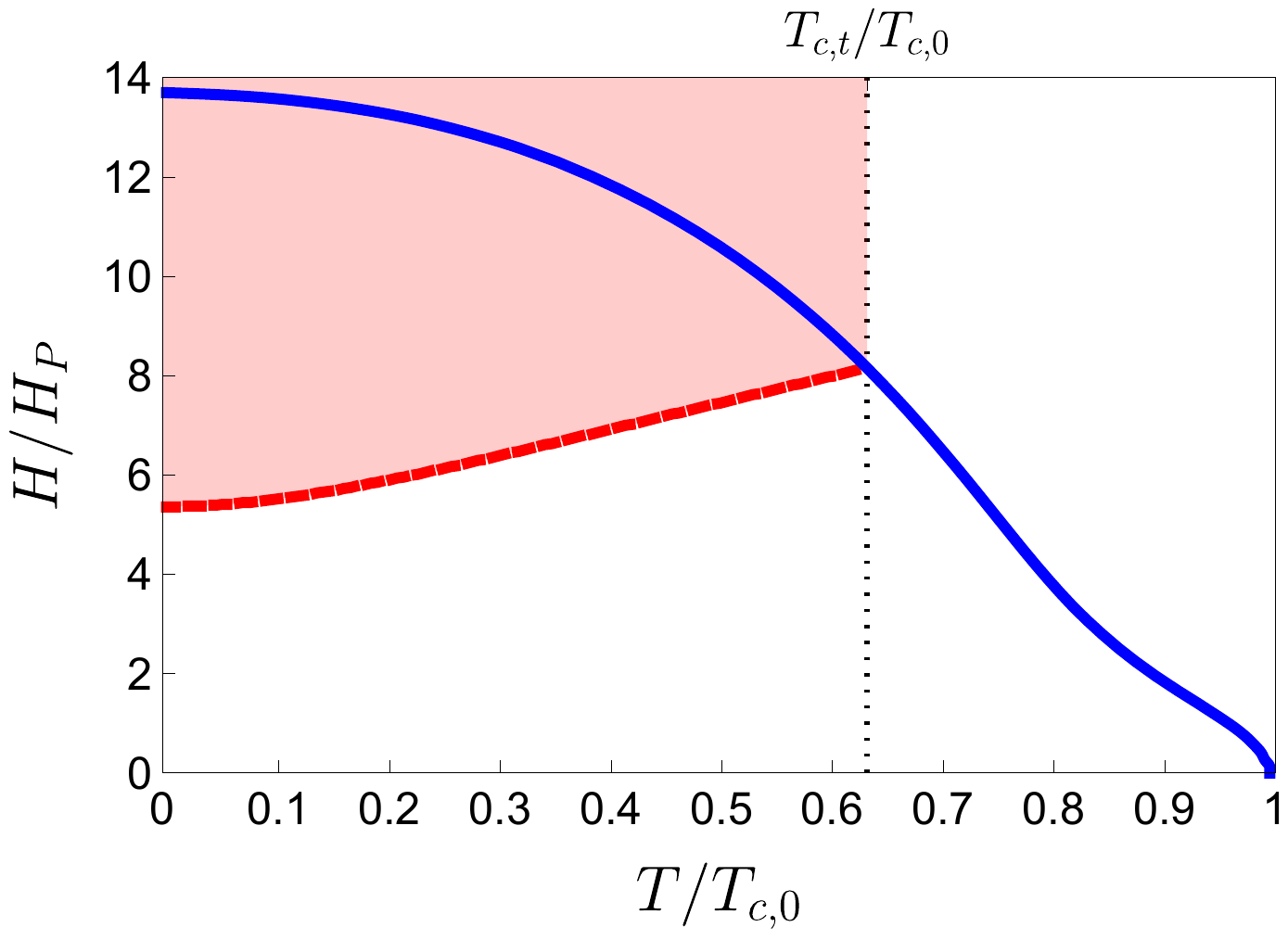}
\put(0,0){(a)}
\end{overpic}
\begin{overpic}[width=0.35\columnwidth]{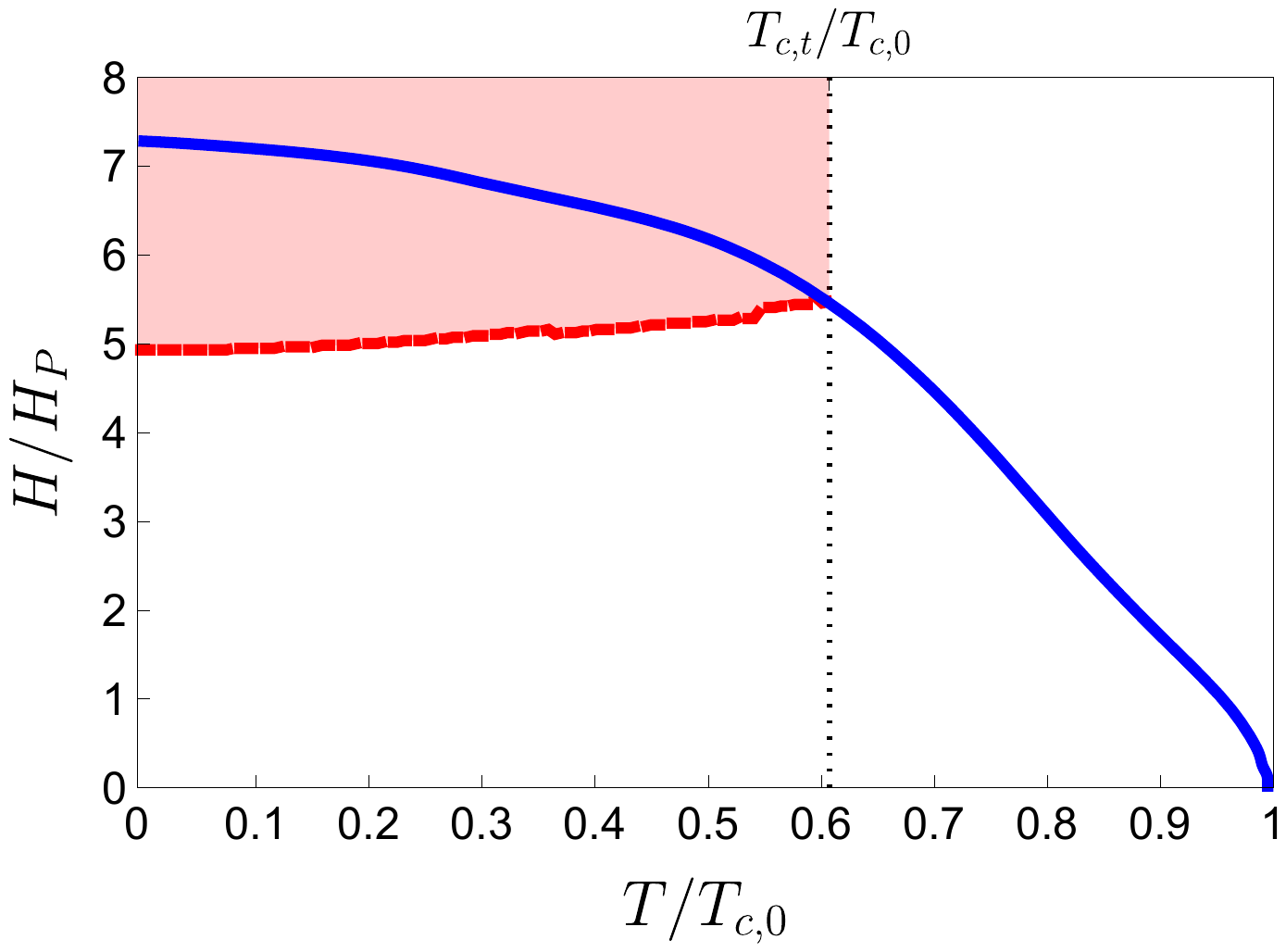}
\put(0,0){(b)}
\end{overpic}
\caption{The phase diagram 
for $\mathcal{N}_M=0.9\mathcal{N}_0$, $g_{\Gamma}=0.9 g_0$ and $\text{max}(g_{M})=0.1g_0$ with (a) $d_{x^2-y^2}$-wave and (b) $d_{xy}$-wave form factors. The blue line gives the upper critical field
of the uniform state, and the red dashed line gives the boundary of
the staggered state (shaded). Here, $T_{c,0}$ is the critical temperature for the $d$-wave state.}\label{fig:appendix_d-waves}
\end{figure}

As a final demonstration of the generality of the phase diagram of the
two pocket model, in Fig.~S\ref{fig:appendix_d-waves} we provide
results for an M-A and $\Gamma$-Z model but with $d$-wave form factors
$f_{\bm k}$. 

\end{document}